\documentclass[fleqn,usenatbib]{mnras}

\usepackage{newtxtext,newtxmath}

\usepackage[T1]{fontenc}

\DeclareRobustCommand{\VAN}[3]{#2}
\let\VANthebibliography\thebibliography
\def\thebibliography{\DeclareRobustCommand{\VAN}[3]{##3}\VANthebibliography}

\usepackage{graphicx}	

\title[Tomographic AP method with redshift errors]{Tomographic Alcock-Paczynski Method with Redshift Errors}

\author[Xiao, Huang, Zheng, et al.]{
Liang Xiao$^{1,2}$,
Zhiqi Huang$^{1,2}$,
Yi Zheng$^{1,2,3}$,
Xin Wang$^{1,2}$,
Xiao-Dong Li$^{1,2}$
\thanks{E-mail: lixiaod25@mail.sysu.edu.cn; wangxin35@mail.sysu.edu.cn; huangzhq25@mail.sysu.edu.cn}
\\
$^{1}$School of Physics and Astronomy, Sun Yat-Sen University, Guangzhou 510297, P.R.China\\
$^{2}$ CSST Science Center for the Guangdong-Hong kong-Macau Greater Bay Area, SYSU \\
$^{3}$Key Laboratory for Particle Astrophysics and Cosmology (MOE)/Shanghai Key Laboratory for Particle Physics and Cosmology, China}

\date{Accepted 2022 October 10. Received 2022 October 08; in original form 2022 June 09}

\pubyear{2021}

\begin{document}
\label{firstpage}
\pagerange{\pageref{firstpage}--\pageref{lastpage}}
\maketitle

\begin{abstract}
The tomographic Alcock-Paczynski (AP) method is a promising method that uses the redshift evolution of the anisotropic clustering in redshift space to calibrate cosmology. It extends the applicable range of AP method to substantially nonlinear scales, yielding very tight cosmological constraints. For future stage-IV slitless spectroscopic surveys, the non-negligible redshift errors might reduce the advantage of the tomographic AP method by suppressing  the resolution of the nonlinear structure along the line of sight. The present work studies how redshift errors propagate to cosmological parameters in the tomographic AP analysis. We use a formula $\sigma_z = \sigma(1+z)^{\alpha} $ to model the redshift errors, with $\sigma$ varying from 0.001 to 0.006 and $\alpha$ varying from 0.5 to 1.5. The redshift errors produce a signal of anisotropic clustering that is similar to a strong finger-of-god effect, which smears out both the AP signal and the contamination caused by the redshift space distortions (RSD). For the target precision of the Chinese Space Station Telescope optical survey ($\sigma\lesssim 0.002$), the decrement of constraining power on the dark energy equation of state is mild ($\lesssim 50\%$), and the suppression of RSD contamination leads to a smaller bias-to-signal ratio. Our results indicate that the tomographic AP method will remain a useful and complementary tool for analyses of future slitless spectroscopic surveys. 

\end{abstract}




\section{Introduction}

The large-scale structure (LSS) of the universe encodes 
an enormous amount of information about the cosmic expansion and structure formation histories.
In the past two decades,
LSS surveys greatly enriched our knowledge and understandings about the Universe 
\citep{2df:Colless:2003wz,beutler20116df,blake2011wigglez,blake2011wigglezb,york2000sloan,Eisenstein:2005su,Percival:2007yw,
anderson2012clustering,alam2017clustering}.
In the next decades, the stage-IV surveys such as 
the Dark Energy Spectroscopic Instrument (DESI)~\footnote{https://desi.lbl.gov/},
the Vera C. Rubin Observatory (LSST)~\footnote{https://www.lsst.org/},
the Euclid satellite~\footnote{http://sci.esa.int/euclid/},
the Roman Space Telescope~\footnote{https://roman.gsfc.nasa.gov/},
and the Chinese Space Station Telescope (CSST)~\footnote{http://nao.cas.cn/csst/}
will measure 
the $z\lesssim2$ Universe to an unprecedented precision,
shedding light on the dark energy problem~\citep{riess1998observational,perlmutter1999measurements,
weinberg1989cosmological,miao2011dark,weinberg2013observational}
and many other fundamental problems.

The Alcock-Paczynski (AP) test~\citep{AP1979} is a pure geometric probe of the cosmic expansion history based on a comparison of measured radial and tangential diameters of known isotropic objects.
The radial and tangential diameters of some distant objects or structures assume the forms of $\Delta r_{\parallel} = \frac{c}{H(z)}\Delta z$ and $\Delta r_{\bot}=(1+z)D_A(z)\Delta \theta$ under a specific cosmological model. Here $\Delta z$ and $\Delta \theta$ are their measured redshift span and angular size, respectively; $D_A$ and $H$ are the angular diameter distance and the Hubble expansion rate derived from the assumed cosmology. With an incorrect theoretical model, i.e., wrong $D_A(z)$ and $H(z)$ functions, the deviation of $\Delta r_{\parallel}$ and $\Delta r_{\bot}$ will cause geometric distortions along the line-of-sight (LOS) and tangential directions. Such distortion can be identified from two-dimensional statistics of galaxies surveys. This method has been successfully employed in many galaxy surveys to provide stringent cosmological constraints~\citep{ryden1995measuring,ballinger1996measuring,matsubara1996cosmological,outram20042df,blake2011wigglez,lavaux2012precision,alam2017clustering, Qingqing2016,KR2018}.

One of the obstacles of using AP on nonlinear scales, where rich cosmological information is encoded, is that the AP signals are polluted by the clustering anisotropies created by the redshift space distortions (RSD). The problem was then resolved by the tomographic AP method~\citep{LI14,LI15}. The key element of tomographic AP is to use the {\it redshift evolution} of clustering anisotropies, which is sensitive to the AP effect but relatively insensitive to the anisotropies created by RSD. This makes it possible to differentiate the AP distortion from the RSD contamination on substantially nonlinear scales of $6$-$40h^{-1}$Mpc.
The addition of rich cosmological information on nonlinear scales makes tomographic AP a very competitive tool of constraining cosmological parameters.
\cite{LI16} firstly applied the method to the 
Sloan Digital Sky Survey (SDSS) Baryon Oscillation Spectroscopic Survey Data Release 12 (BOSS DR12, https://www.sdss.org/dr12) galaxies, and achieved $\sim40\%$ improvements in the constraints on the dark matter abundance $\Omega_m$ and dark energy equation of state (EOS) $w$, when combining the method with the datasets of 
the Planck measurements of cosmic microwave background~\citep{planck2016ade}, 
type Ia supernovae~\citep{betoule2014}, baryon acoustic oscillations (BAO)~\citep{Anderson2014}
and local $H_0$ measurement~\citep{Riess2011,Efstathiou2014}.
In follow-up studies, \cite{LI18,Zhang2019} studied the constraints on models with possible time-evolution of dark energy EOS, and showed that the method can reduce the errors of the parameters by $\sim50$\%. \cite{LI19} forecast the performance of the tomographic AP method on future DESI data and discovered that DESI+CMB constraints on dynamical dark energy models can be improved by a factor of $\sim 10$.

The tomographic AP method is so far the best method for isolating the AP signal from the RSD effects and conduct AP test on scales $\lesssim 40h^{-1}\ \rm Mpc$.
Thus, the systematics of the method, mainly coming from the redshift dependence of the anistropy produced by the RSD, should be carefully evaluated, so as to ensure that the analysis is robust and to make good preparations for the application of the method to the future LSS experiments. As the first work applying the method to real observational data, \cite{LI16} utilized the Horizon Run 4 (HR4) N-body simulation~\citep{kim2015horizon} to test and estimate the systematics, and claimed that its effect on the final cosmological results is insignificant. Furthermore, in the follow-up work~\citep{LI18}, the authors found that the systematics can only lead to $\approx0.3\sigma$ shift in the derived cosmological constraints. These works manifested that, for stage-III surveys, the systematics of the method is less significant compared with the statistical error of the survey itself.

In the next decade, the commission of stage-IV surveys will bring greater challenge to the calibration of the systematics. These surveys will significantly reduce statistical error, observe structures at higher redshift with more RSD contamination, and extract more cosmological information from non-linear scales. Thus, more research is required to ensure that the tomographic AP method can be safely applied to these surveys. The authors of~\cite{Park:2019mvn} studied the the systematics based on the cosmological simulations conducted in five distinct cosmologies, and found that cosmological dependence of the systematics can produce visible changes in the derived cosmological constraints.
\cite{Ma_2020} found that systematics becomes more significant at $z\sim 1$ compared with its behavior at lower redshift, and advocated that one can use fast simulations to efficiently estimate the systematics in order to lower the computational cost of systematics estimation for these huge surveys. 

Despite the existence of the aforementioned works, none of these studies discusses the systematics caused by redshift uncertainties of the galaxies. In the next few years, slitless instruments will be used in the Chinese Space Station Telescope~\citep[CSST]{Gong_2019} and the Euclid satellite~\citep{EUCLID} to determine the spectroscopic redshifts of galaxies.
While this methodology significantly improves the convenience and efficiency of the spectroscopic survey, the cost is that, compared with the traditional fiber-based surveys, the errors of the measured redshifts of the galaxies will increase by one order of magnitude.
The redshift errors are expected to distort the estimated comoving distances of the galaxies, smear the clustering information below the scale of this distortion, and therefore produce a signal of anisotropy similar to the finger-of-god (FOG) effect.
This provides a new source of systematics for the tomographic AP analysis, whose potential impact is to be investigated in the present work. Without loss of generality, we constrain energy density of mass $\Omega_m$ and EOS of dark energy $w$ in this paper. This work is especially aimed at the CSST (Chinese Space Station Telescope), however, our work is also useful for other spectroscopic slitless surveys.

This paper is organized as follows.
In Section 2, we introduce the simulation materials used in the paper. 
In Section 3, we firstly review the methodology of the tomographic AP method, and then describe how we consider the redshift errors in the analysis.
In Section 4, we present our results, illustrating how the redshift errors affect the systematic bias and statistical error of the tomographic AP method.
We conclude in Section 5.

\section{Data}

We use the Big MultiDark Planck (BigMD) simulation as the mock data to test the effect of redshift errors on the tomographic AP method. As one of the largest simulation
among the series of MultiDark simulations,
the BigMD simulation was produced with $3\,840^3$ particles in a cubic box with side length $(2.5h^{-1}\rm Gpc)^3$,
assuming a $\Lambda$CDM cosmology with Planck parameters
$\Omega_m = 0.307115$, $\Omega_b = 0.048206$, $\sigma_8 = 0.8288$, $n_s = 0.9611$, 
and $H_0 = 67.77\ {\rm km}\ s^{-1} {\rm Mpc}^{-1}$~\citep{BD}. The initial condition of the simulation was set by using the Zeldovich approximation~\citep{Crocce2LPT} at redshift $z_{\rm init} = 100$.

Throughout this analysis, we use the halos and subhalos identified by the ROCKSTAR halo finder~\citep{ROCKSTAR} at snapshots $z=0.607$ and $z=1$. Note that it is necessary to use lightcone to model systematics when dealing with real observational data ~\citep{LI16}. In this work, the main issue is the effect of redshift errors, so we use snapshots for simplicity .

We investigate the difference between the anisotropic correlation function measured at the two snapshots to reveal the redshift evolution of the anisotropic clustering. 
A constant number density $\bar n= $ 0.001 $(h^{-1}\rm Mpc)^{-3}$ is imposed by applying a cut on the mass of the objects.

\section{Methods}


\subsection{The tomographic Alcock-Pazynski method}

The AP effect arises when an incorrect 
cosmological model is used to calculate
the distances of the observed objects,
whose comoving size takes the form of
\begin{equation}
    \Delta r_{\parallel} =\frac{c}{H(z)}\Delta z,\ \Delta r_{\perp} = (1+z)D_A(z)\Delta \theta.
\end{equation}
In case of a flat universe composed by 
a dark matter component taking a fraction of $\Omega_m$ 
and a dark energy component with a constant EOS $w$, 
the Hubble parameter and the angular diameter distance take the forms of
\begin{eqnarray}
    H(z)&=&H_0 \sqrt{\Omega_m a^{-3}+(1-\Omega_m)a^{-3(1+w)}},\ \\
    D_A(z)&=&\frac{c}{1+z}r(z)=\frac{c}{1+z}\int_0^z \frac{dz^\prime}{H(z^\prime)},
\end{eqnarray}
where $a=1/(1+z)$ is the cosmic scale factor,
$H_0$ is the present value of Hubble parameter, 
and $r(z)$ is the comoving distance.
For simplicity, we characterize the anisotropy 
by the integrated two-point correlation function,
\begin{equation}
    \xi_{\Delta s}(\mu) \equiv \int_{s_{\rm min}}^{s_{\rm max}} \xi(s,\mu) ds, \label{equ:integral_s}
\end{equation}
where $s$ is the distance between the two points, and $0\le \mu\le 1$ is cosine of the angle between the line of sight (LOS) and the line connecting the two points. The integral bounds are taken to be $s_{\rm min}=6h^{-1}\mathrm{Mpc}$ and $s_{\rm max}=40h^{-1}\mathrm{Mpc}$. Following our previous works, we do not use the $>40h^{-1}\mathrm{Mpc}$, due to three reasons: 
\begin{itemize}
    \item The traditional anistropic BAO analysis already covers the AP effect on th BAO scale, so to avoid overlapping information, we do not include that region in our analysis \footnote{See Figure B1 of ~\cite{Zhang:2018jfu}; besides, Figure A1 of ~\cite{Zhang:2018jfu} shows our method is almost un-correlated with traditional anistropic BAO analysis.}
    \item The uniqueness of our method lies in that it can cover the relatively non-linear clustering scale not covered by most common methods. So we focus on the non-linear clustering scales, making our method complementary to traditional methods.
    \item We use $\xi_{\Delta s}$ as the statistical quantity. On large scales, the error of $\xi$ becomes rather large. Including this region actually increases the total error of $\xi_{\Delta s}$ and worsen the results (Figure 12 of ~\cite{LI16} shows that setting $s_{\rm max}=50h^{-1}\mathrm{Mpc}$ slightly weakens the constraint).
\end{itemize}
In this work we will not check the robustness of the method against the choices of $s_{\rm min}$ and $s_{\rm max}$. ~\cite{LI16} checked that the cosmological constraints maintain statistically consistent in $ 1\sigma $ if one used even smaller $ s_{\rm min} = 2,\ 4 h^{-1} \rm Mpc $ or larger $ s_{\rm max} = 50 h^{-1} \rm Mpc $ \footnote{For checking the robustness of this method it is worthy trying larger $ s_{\rm max} $, since on large clustering scales the systematics is lower and the clustering physics is better understood. We leave this study to future works.}.

To mitigate the systematic uncertainty caused by galaxy bias and the clustering strength, a further normalization is imposed
\begin{equation}
\hat \xi_{\Delta s}(\mu) \equiv \frac{\xi_{\Delta s}(\mu)}{\int_0^{\mu_{\rm max}} \xi_{\Delta s}(\mu)d\mu}.
\end{equation}
In this case, we use a cut $\mu < \mu_{\rm max}$ to eliminate the non-linear FoG effect and fiber collisions that are more intense in the LOS direction ($\mu\rightarrow1$). The redshift dependence of the anisotropy is thus quantified via
\begin{equation}\label{Eq:deltaxi}
    \delta \hat \xi_{\Delta s}(z_1, z_2, \mu) \equiv \hat \xi_{\Delta s}(z_2, \mu)-\hat \xi_{\Delta s}(z_1, \mu),
\end{equation}
which is the difference in anisotropy between $z_1$ and $z_2$ redshifts. The statistical significance of this variable is then quantified using the $\chi^2$ function, defined as
\begin{equation}
    \chi^2 ={\bf p}^{\mathrm{T}} \cdot {\bf Cov}^{-1} \cdot {\bf p}. \label{equ:chi2}
\end{equation}
Here ${\bf p}$ is the redshift evolution of the anisotropy caused by AP, which we will define in a moment.  The statistical uncertainty of $\bf p$ is quantified by the the covariance matrix $\bf Cov$, which is estimated by 512 subsamples. The details are in Sec. \ref{section 4.2}. We are not going to apply Hartlap coefficient~\citep{Hartlap:2006kj} because it's a const when keeping a constant number of $ \mu $ bins and thus will not affect the ratio of two contour areas (with and without redshift error).

Effects such as the RSD also contribute non-zero
redshift evolution of anisotropy to $\delta \hat \xi_{\Delta s}(z_1, z_2, \mu)$.
To correct these systematics\footnote{The RSD effect is the major contribution of the systematics, but there also exists other sources of systematics, mainly comes from observational effects such as the seletion bias, the fiber collision, and so on. When dealing with obsevational data we need to consider and include them in the mock surveys (see the analysis procedure of ~\cite{LI16}).}, we define ${\bf p}$ as
\begin{equation}
    {\bf p}(z_i,z_{i+1},\mu_j) \equiv \delta \hat \xi_{\Delta s}(z_i, z_{i+1},\mu_j) - \delta \hat \xi_{\Delta s, {\rm sys}}(z_i, z_{i+1},\mu_j), \label{equ:pz}
\end{equation}
where discretization of $\mu$ is also applied. In particular, we use 19 $\mu$ bins in the present work.
We follow~\cite{LI15,LI16}, where the systematics $\hat \xi_{\Delta s, {\rm sys}}$ and the covariane matrix $\mathbf{Cov}$ are evaluated by mock survey data from N-body simulations.

\subsection{Modelling the redshift errors}

Although being successful when applied to SDSS galaxies, the tomographic AP method faces several challenges if it is to be applied to future stage-IV surveys. The increased survey volume and precision reduces the statistical uncertainty, necessitating a more precise systematic reduction in data analysis. While the challenges brought by the RSD have been discussed in detail in previous works~\citep{Pan:2019vky,Ma_2020}, in this work we study the effect caused by the redshift errors.  

In the slitless surveys, the measured redshifts of the galaxies have larger errors compared with traditional fiber based surveys. Taking the CSST optical survey as an example, its redshift errors are estimated to be~\citep{Gong_2019} 
\begin{equation}
\sigma_z\approx0.002 (1+z), \label{equ: CSST_error}
\end{equation}
which is an order of magnitude larger than the redshift errors of current spectroscopic surveys. The redshift errors create distortions of order a few $\mathrm{Mpc}$ in the calculated comoving distance of the objects, potentially contaminating the tomographic AP signals.

The influence of the redshift errors is manifested in two aspects. Firstly, the above redshift errors are comparable to a strong FOG effect with peculiar velocity $v\sim600{\rm km/s}$. This results in a massive amount of fictitious anisotropy in the observed galaxy sample, and will significantly alters the shape and amplitude of  $\hat\xi_{\Delta s}$. Such a strong effect may manifest as significant systematics in the analysis procedure, resulting in bias in the estimation of cosmological parameters. Secondly, large redshift errors can produce a significant smearing effect. This reduces statistical power of the tomographic AP method, hence weakening the cosmological constraints.

In general, redshift dependence more sophisticated than Eq.~\ref{equ: CSST_error} is often observed in reality, due to the complications related with the instrumentation, the method used to estimate the redshift, the properties of the objects (e.g. emission lines, luminosity), and the selection bias. To give some concrete examples, we parameterize the redshift errors as
\begin{equation}\label{eq:error_scheme}
    \sigma_z = \sigma (1+z)^\alpha
\end{equation}
where $\sigma$ and $\alpha$ are constants. Because we only use two tomography bins, the complications of $z$-dependence can be effectively characterized by the $\alpha$ parameter. Besides, in surveys like CSST the redshift error could be large at $ z>1 $(due to the lack of Ly-$\alpha$ lines at high redshift), so we adopt different power-law formulas to make a relatively general discussion. It is difficult to use more detailed formular since 1) details about the redshift error of CSST is so far unknown, and 2) more complicated formular will make the analysis too complicated and losing generality.

We will examine eight redshift error schemes, divided into two sets. In one set, we fix $\alpha=1$ and choose $\sigma = 0.001, 0.002, 0.004, 0.006 $. In the other set, we fix $\sigma=0.002$, and consider $\alpha = 0.50, 0.90, 1.10$ and $1.50$, to investigate the non-standard redshift dependence of the error.

The redshift errors affect the cosmological analysis by distorting the estimated comoving distances of the objects. The distortion is characterized by a scale
\begin{equation}
    \sigma_r = \frac{c\sigma_z}{H(z)}.
\end{equation}
Information of galaxy clustering on scales $\lesssim \sigma_r$ is smeared out by the redshift uncertainties.
\begin{figure*}
    \centering
    \includegraphics[width=0.8\textwidth]{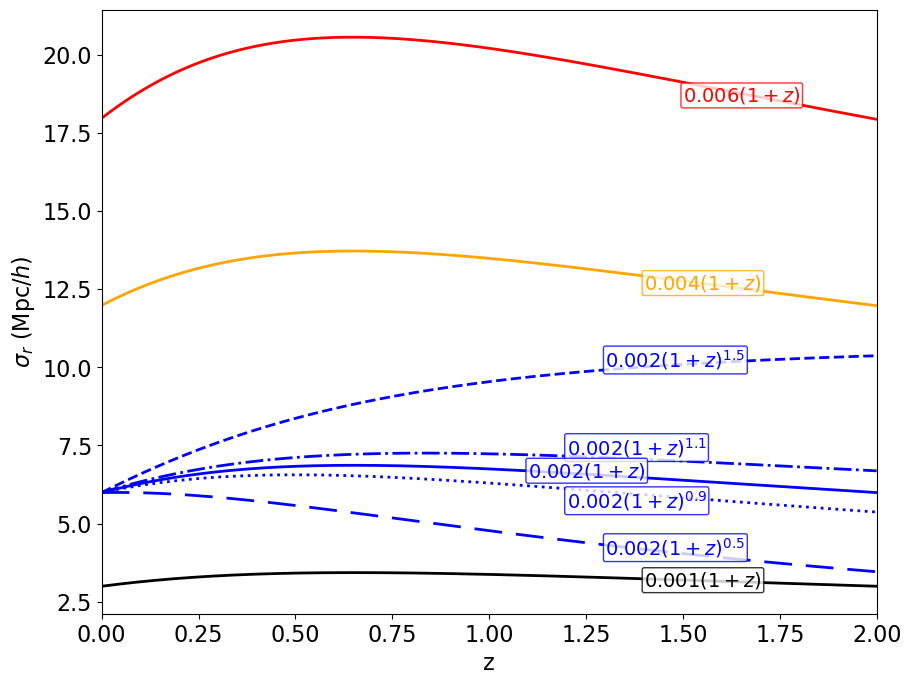}
    \caption{The distortion in the comoving distance caused by redshift errors, according to Eq. \ref{eq:error_scheme}.}
    \label{figure:sigma_r}
\end{figure*}
As shown in Fig.~\ref{figure:sigma_r}, comoving distance distortions caused by various redshift error schemes have quite different amplitudes and redshift dependence. Given that the tomographic AP method typically utilizes information of clustering on scales of $6$-$40 h^{-1}{\rm Mpc}$, we expect non-negligible systematics in most schemes. The systematics in the $\sigma=0.004$ and $\sigma=0.006$ schemes are expected to be quite significant.

\section{Analysis and Results}

To investigate the effect of redshift errors, we simulate them by adding Gaussian errors to the comoving distances. In the snapshot data, we choose the third Euclidean coordinate as the LOS direction, and distort it by both the RSD and the redshift errors. To maintain consistency, we refer to this direction as the LOS when measuring the correlation functions during the pair counting procedure. The measured $\hat \xi_{\Delta s}(\mu)$s are used to investigate the systematic bias and statistical error introduced by the redshift errors.

For the two BigMD snapshots at $z_1$ and $z_2$, the simulated observables are
\begin{equation}
\delta \hat\xi_{\Delta s}(z_2, z_1) \equiv \hat\xi_{\Delta s}(z_2) - \hat\xi_{\Delta s}(z_1). \label{equ:zs}
\end{equation}
Although many other redshift slicing schemes are possible~\citep{LI16, LI19}, the particular choice made in the present work, $z_1=0.607$ and $z_2=1$, is representative for stage-IV surveys that typically collect many spectroscopic galaxy samples around $z\sim 1$. For CSST, the slitless redshift errors may blow up at $z\gtrsim 1$, as CSST may not be able to detect the Ly-$\alpha$ lines of $z\gtrsim 1$ galaxies. In addition, \cite{Ma_2020} showed that the systematics of the tomographic AP method is significantly larger for redshift slices $z\gtrsim 1$. Thus, we do not include more tomography slices at higher redshift.

With the above settings, we are able to quantify the {\it signal}, which is the AP component in $\delta \hat\xi_{\Delta s}$, and the {\it bias}, which is the residual $\delta \hat\xi_{\Delta s}$ excluding the AP effect. For the particular choice of redshift slicing in the present work,  the systematic bias is given by
    \begin{equation}
        {\rm  Bias}\equiv \left(\hat\xi_{\Delta s}(\mu)|_{z=1}-
        \hat\xi_{\Delta s}(\mu)|_{z=0.607}\right)_{\rm in\ correct\ cosmology}, \label{equ: Bias}
    \end{equation}
and the AP signal can be written as 
    \begin{equation}
        {\rm  Signal}\equiv \left(\hat\xi_{\Delta s}(\mu)|_{z=1}-
        \hat\xi_{\Delta s}(\mu)|_{z=0.607}\right)_{\rm in\ wrong\ cosmology}
         \ - \ {\rm Bias}. \label{equ: Signal}
    \end{equation}

Compared with the quantification of the bias and the signal, the quantification of the statistical power is more complicated. To estimate that, we divide the entire BigMD sample into $8^3$ subsamples, and estimated the covariance of the signal measured in these subsamples. We compute the values of $\chi^2$ using Eq.~\ref{equ:chi2}, over a sufficiently large range in $\Omega_m$-$w$ space, using various error schemes. In particular, we choose the parameter ranges as  $0.2 \le \Omega_m \le 0.5, -1.5 \le w \le -0.5$, and compute the chi squares on  a uniform $30 \times 30$ grid, so that to derive the constraints on the parameters. 

In the following two subsections, we discuss how the redshift errors affect the systematic bias and statistical power of the method.

\begin{figure*}
    \centering
    \includegraphics[width=0.95 \textwidth]{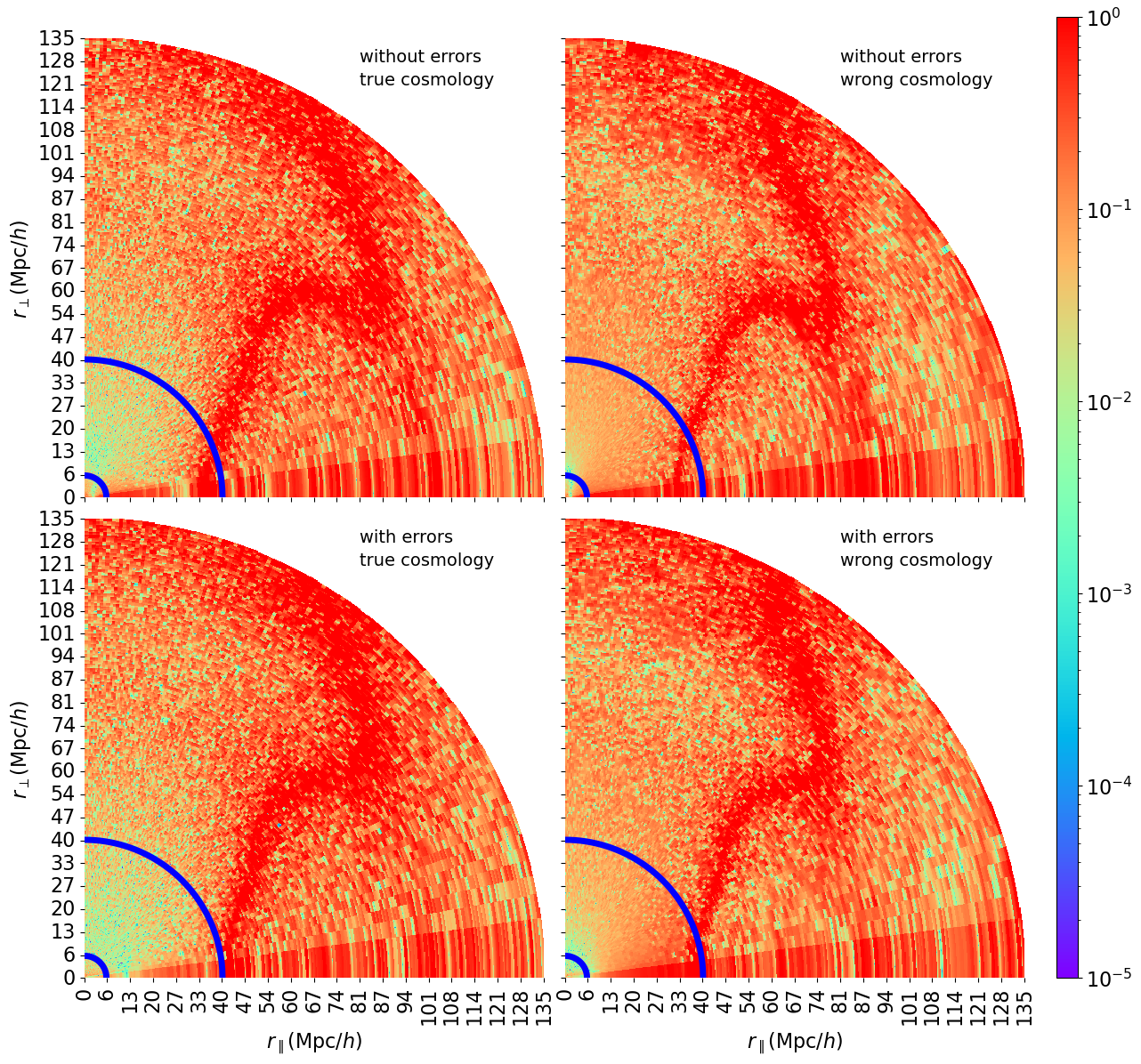}
    \caption{The difference between the two-point correlation functions measured at $z=0.607$ and $z=1$. In the first and second rows, we plot the case without redshift error and the case with redshift error taking a form of $ \sigma_z = 0.002(1+z) $, while in the left and right columns we show the measured redshift evolution when adopting the true anc an incorrect cosmology \protect\footnotemark with $\Omega_m = 0.4071$, respectively. All plots are made in  $ r_\parallel - r_\perp $ space. The clustering region $ r \in [6, 40] \mathrm{Mpc}/h $ is marked by blue lines.}
    \label{figure:r_space}
\end{figure*} \footnotetext{Here we compare the signal and noise in the true and wrong cosmologies. In real analysis, one does not know the true cosmology. He/She seeks for the true cosmology by seeking for parameters that can minimize the signal of AP (quantified by the redshift evolution of anisotropy).}

\begin{figure*}
    \centering
    \includegraphics[width=0.95 \textwidth]{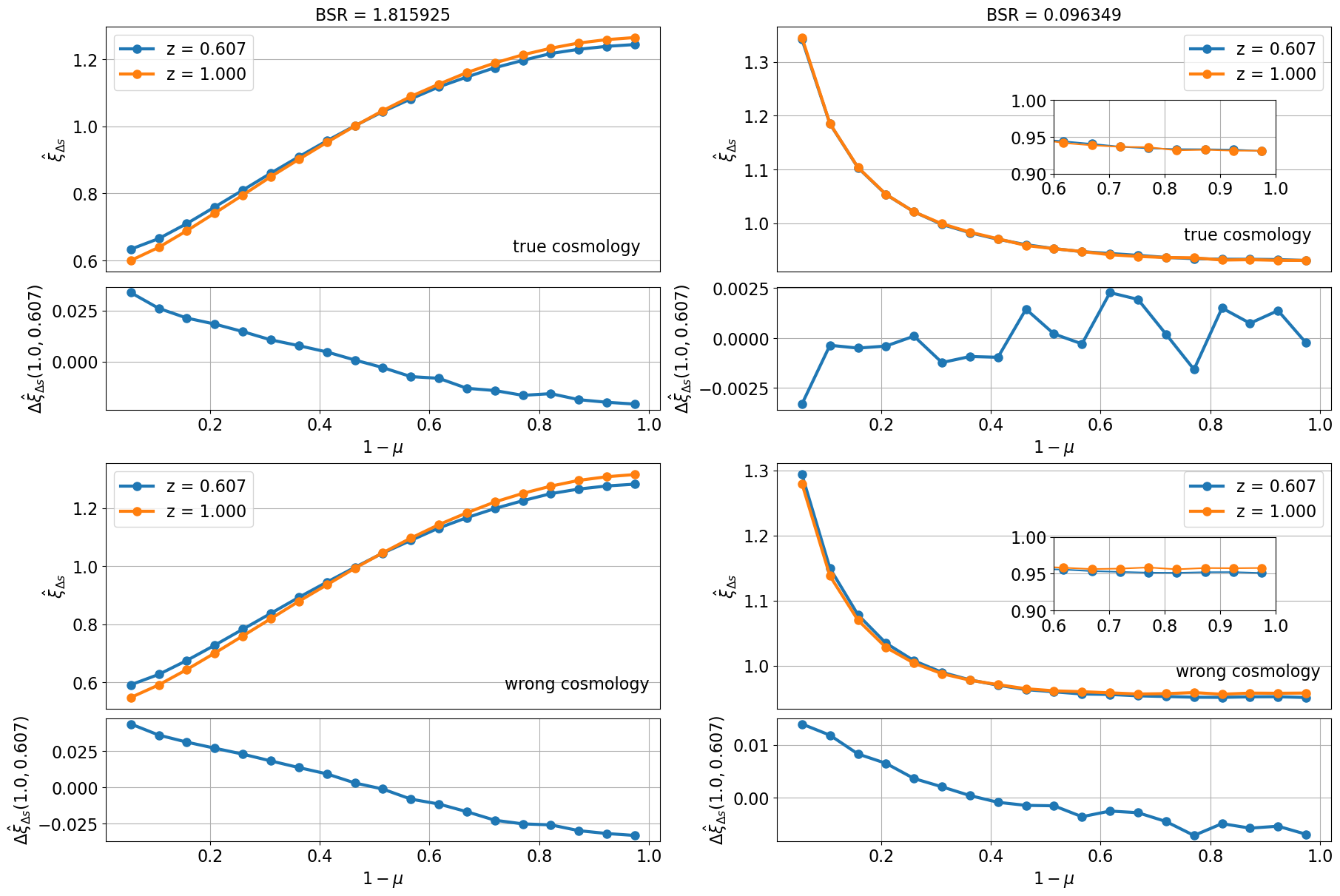}
    \caption{Comparison of integrated two-point correlation function with and without redshift errors. An incorrect  cosmologies ($\Omega_m = 0.4071$) is used in the lower panels to generate the tomographic AP signals.The BSR is calculated using Eq.~\ref{equ:BSR}, where $\sigma_z \equiv 0 $ for the left panels and $\sigma_z = 0.002(1+z) $ for the right panels are used, respectively. At the bottom of each panel a narrow subfigure illustrates the difference between two redshift snapshots. The zoom-in subfigures within the right panels elucidate some of the details.}
    \label{figure:ximu}
\end{figure*}

\subsection{The systematic bias}

The systematic bias due to the RSD contamination is discussed in detail in~\citep{LI14,LI15,LI16,Ma_2020}. The bias discussed in this work is a combination of the RSD contamination and the redshift errors.

Firstly, as an intuitively illustration, in Figure~\ref{figure:r_space} we plot Eq. \ref{equ: Bias} and Eq. \ref{equ: Signal} in $ r_\parallel $-$ r_\perp $ space. Clearly, in case that one adopted a wrong cosmology (for which we choose $\Omega_m=0.4071$), the redshift evolution tends to become more significant in all clustering scales, and this trend maintains if we consider a redshift error scheme $\sigma_z = 0.002(1+z)$. This figure suggests that, it is feasible to conduct the tomographic Alcock-Paczynski test in case that there exists galaxy redshfit errors.


To see how the redshift errors affect the measurements of the anisotropic clustering, in Figure~\ref{figure:ximu} we show $\hat\xi_{\Delta s}(\mu)$ measured in the correct cosmology (upper panels) and the wrong cosmology with $\Omega_m=0.4071$ (lower panels). 
The left panels show the results in the absence of redshift error, which we hereafter refer to as the no-redshift-error case, and the right panels show the measurements in case of $\sigma_z = 0.002(1+z)$.
Clearly, the redshift errors have a substantial effect on the measurement,
producing a strong tilting of the curve near the LOS (the region near $1-\mu=0$), which can be analogous to a very strong FoG effect.
Besides the sharp peak near the LOS direction, the effect is so strong that, it affects the clustering pattern in most of the directions, and thus modifies the entire shape of $\hat \xi_{\Delta s}(\mu)$.

Despite the big impact on the $\mu$-dependence of $\hat\xi_{\Delta s}$, however, the redshift errors do not introduce too much redshift evolution of $\hat\xi_{\Delta s}$. In case that the cosmological parameters are correct (i.e. no AP distortion), the dominant source of redsfhit evolution of  $\hat\xi_{\Delta s}$ is the RSD contamination. As shown in the upper panels of Figure~\ref{figure:ximu}, when the dominant RSD contamination is smeared out by redshift errors, the difference between the $\hat\xi_{\Delta s}(\mu)$ measured in the two redshifts is $\lesssim0.25\%$,  an order of magnitude smaller than the no-redshift-error case. The lower panels of Figure~\ref{figure:ximu} show that the redshift evolution of $\hat\xi_{\Delta s}$ is more visible when a wrong cosmology is used. This indicates that, compared to the RSD contamination, the tomographic AP signal is typically less suppressed by the redshift errors.

To quantitatively compare the systematics with and without redshift errors, we define the bias-to-signal ratio (BSR) to measure the harm of the systematics,
\begin{equation}
    {\rm BSR}\equiv \left|\frac{\rm Bias}{\rm Signal}\right|, \label{equ:BSR}
\end{equation}
which is simply the ratio between the systematic bias and the cosmological signal.
In practice, the ratio between the two functions of $\mu$ may not be elucidating. 
To simplify the problem, we imposed an integration\footnote{The only purpose of Eq.~\ref{eq:transform} is to define a quantity which is convenient for us to define a quantity characterizing the Bias, the Signal, and the BSR (originally the Bias and Signal are all curves, which are difficult for quantification purpose). The choice of 0.4 is a bit arbitrary. Although rough, the defined quantity is enough for the purpose of our work.} 
\begin{equation}\label{eq:transform}
\int_{\mu=0}^{\mu=0.4} d\mu\ \hat\xi_{\Delta s}(\mu),
\end{equation}
to convert each function to a single number, and define the BSR as the ratio between the two numbers.

Eq.~\ref{eq:transform} is an integration conducted in the near-LOS region. It is an effective description of the tilting of the curve near the LOS region. Since the dominant phenomenon of all effects discussed in this work (the AP distortion, the RSD, the redshift errors) is the tilting of $\hat\xi_{\Delta s}(\mu)$, Eq.~\ref{eq:transform} is a reasonable simplification that captures the major phenomenon of these effects.

\begin{figure*}
    \centering
    \includegraphics[width= 1.0\textwidth]{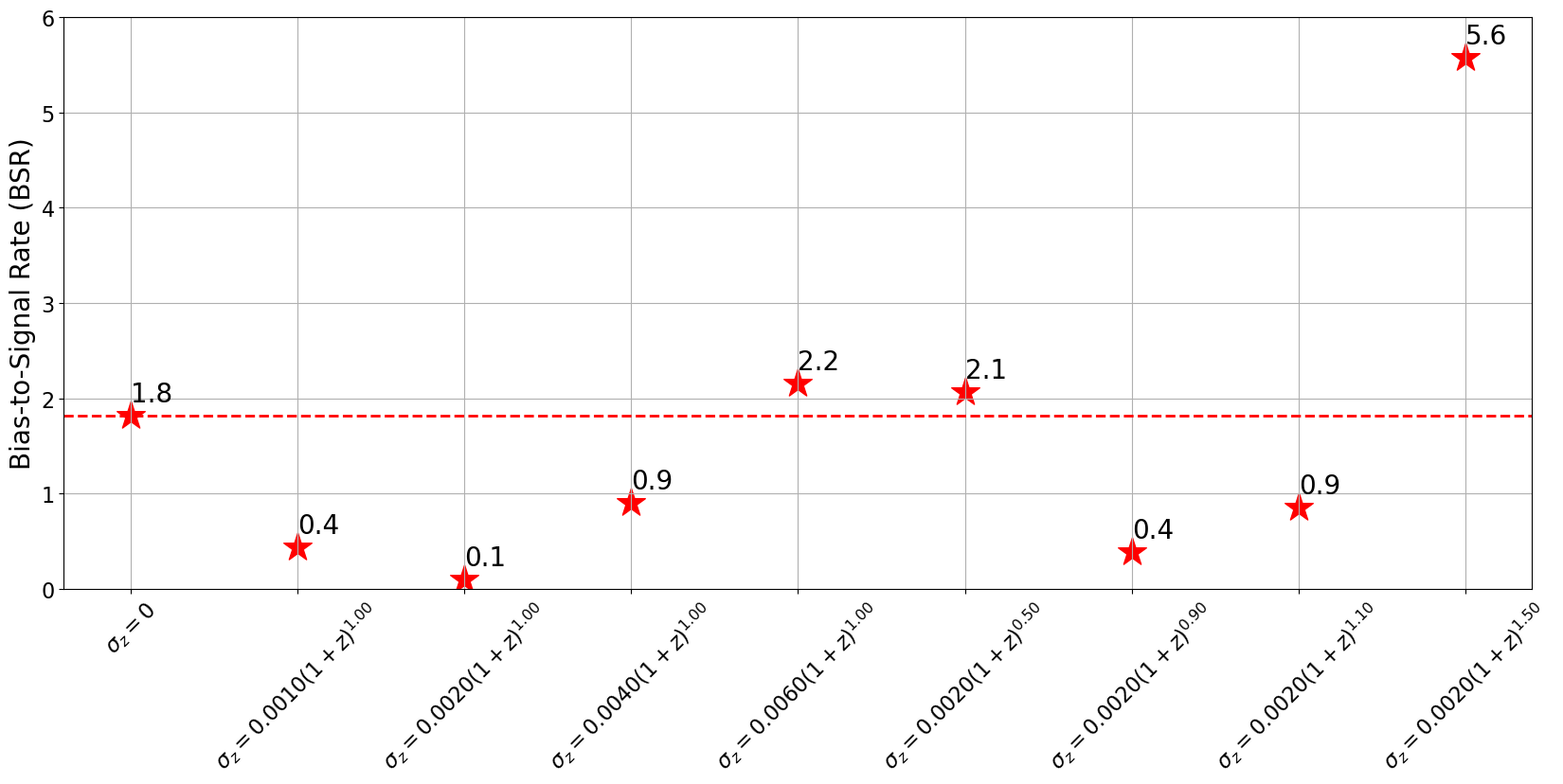}
    \caption{The BSR values (defined in Eq.~\ref{equ:BSR}) of 8 redshfit error schemes in the vicinity of Eq.~\ref{equ: CSST_error}. 
    The dashed red line represents the BSR in case of no redshift error. 
    In most cases, the redshift errors do not significantly increase the BSR. }
    \label{figure:BSR}
\end{figure*}

As shown in Figure~\ref{figure:BSR}, in most of the eight schemes of redshift errors, the BSR is lower than the no-redshift-error case. For the linear redshift error ($\alpha=1$) with $\sigma=0.001$, $0.002$, $0.004$, we find a BSR of $0.441$, $0.096$, $0.904$, which are $76\%$, $95\%$, $50\%$  smaller than the no-redshift-error BSR $1.8$, respectively. In the case of $\sigma=0.006$, we find a worse BSR $2.153$ that is $19\%$ larger than the no-redshift-error case.  In non-linear error ($\alpha\ne 1$) schemes, the $ \alpha= 0.5$, $0.9$, $1.1$, $1.5$ cases lead to BSR rates of $2.1$, $0.39$, $0.85$ and $5.6$, respectively. Thus, the BSR is only enhanced unless there exists very large redshift errors (e.g. $\sigma=0.006$) or a large deviation from linearity (e.g. $\alpha = 1.50 $). If the amplitude and non-linear redshift dependence of the redshift error is not too large, we always find a suppressed BSR.

We try to explain why in some specific schemes, the redshift errors bring suppressed BSRs. The total amount of systematic bias is a combination of the RSD and the redshit errors,
\begin{equation}
    \sigma_{\rm total} =\sqrt{\sigma_{\rm RSD}^2 + \sigma_{z}^2}.
\end{equation}
where $ \sigma_{\rm RSD} $ is a simplification of the bias caused by RSD. In case of $\sigma_z \gg \sigma_{\rm RSD}$ we can approximate it as 
\begin{equation}\label{eq:sigRSD}
    \sigma_{\rm total}  \approx  \sigma_z + \frac{\sigma_{\rm RSD}}{2\sigma_z^2}  \sigma_{\rm RSD}, 
\end{equation}
the contribution of $\sigma_{\rm RSD}$ is therefore suppressed by a factor of $\frac{\sigma_{\rm RSD}}{2\sigma_z^2}$.
In case that $\sigma_z$ does not introduce a significant redshift evolution in $\sigma_r$, the redshift evolution of $\sigma_r$ mainly comes from the RSD, and a large $\sigma_r$ can suppress the contribution of the RSD, leading to a more redshift invariant $\sigma_r$ \footnote{As can be seen in Figure~\ref{figure:sigma_r}, the $\sigma_r$ produced by many redshift error schemes do not have significant redshift dependence, so they can play a role of suppressing the RSD and ``stablizing'' the behaviour of $\sigma_r$ in a wide range of redshift.}.

In the above analysis, however, we also see that $\sigma_z=0.002(1+z)^{1.5}$ produces a very high BSR, which is three times larger than that of the no-redshift-error case. Figure~\ref{figure:sigma_r} implies that the underlying reason is the huge redshift dependence of $\sigma_r$ created by this redshift error. Here we emphasize that, although in such cases the systematics can severely affect the cosmological analysis, as long as we have accurate knowledge about the redshift errors, this large effect can still be quantified and corrected in the analysis. In other words, a large BSR does not necessarily mean that the analysis is not feasible, rather the feasibility of the analysis more depends on how accurate we can model the systematics and calibrate it.


\subsection{The statistic power}\label{section 4.2}

In this subsection, we proceed to investigate how the redshift errors affect the statistical power of the tomographic AP method. We split each BigMD snapshot sample into $8^3$ subsamples, each having a boxsize $(312.5 h^{-1}\rm Mpc)^3$, which is still large enough considering that our statistics is conducted on clustering scales of $6$-$40 h^{-1} \rm Mpc$. Then we compute the covariance of the correlation function with these 512 subsamples.

\begin{figure*}
    \centering
    \includegraphics[width= 0.9\textwidth]{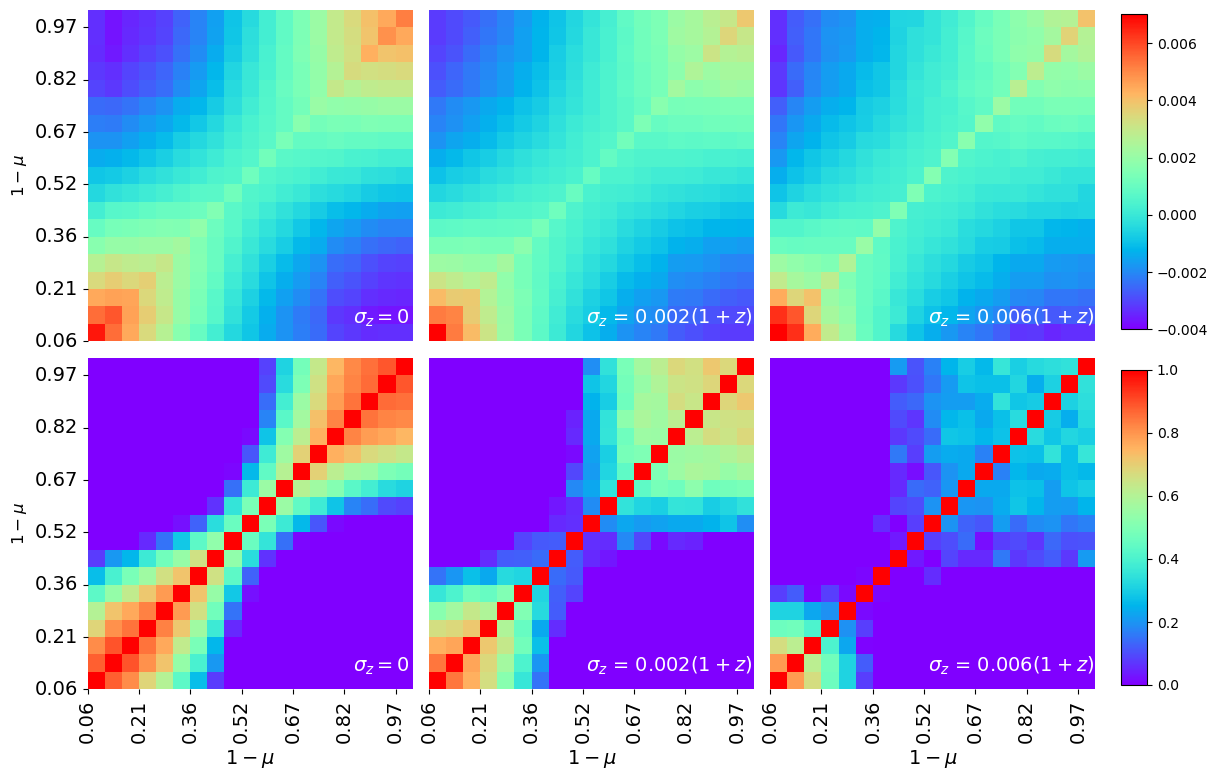}
    \caption{Covariance matrices (upper panels) and correlation coefficients (lower panels) of $\hat\xi_{\Delta s}(\mu,z=1)-\hat\xi_{\Delta s}(\mu,z=0)$. 
    }
    \label{fig:cov}
\end{figure*}

In Figure~\ref{fig:cov}, we show for different redshift-error schemes the covariance matrices (upper panels) as well as the correlation coefficients (lower panels). In case of $\sigma_z=0.002(1+z)$, compared to the no-redshift-error case, the values of the the covariance matrices are only mildly affected, and the correlation among different $\mu$-bins is slightly suppressed. When a larger redshift error $\sigma_z=0.006(1+z)$ is assumed, the correlation matrix is more significantly suppressed. Thus, redshift errors have non-negligible effect on the statistical properties of the tomographic AP signals, and may alter the derived cosmological constraints.

\begin{figure*}
    \centering
    \includegraphics[width= 0.9\textwidth]{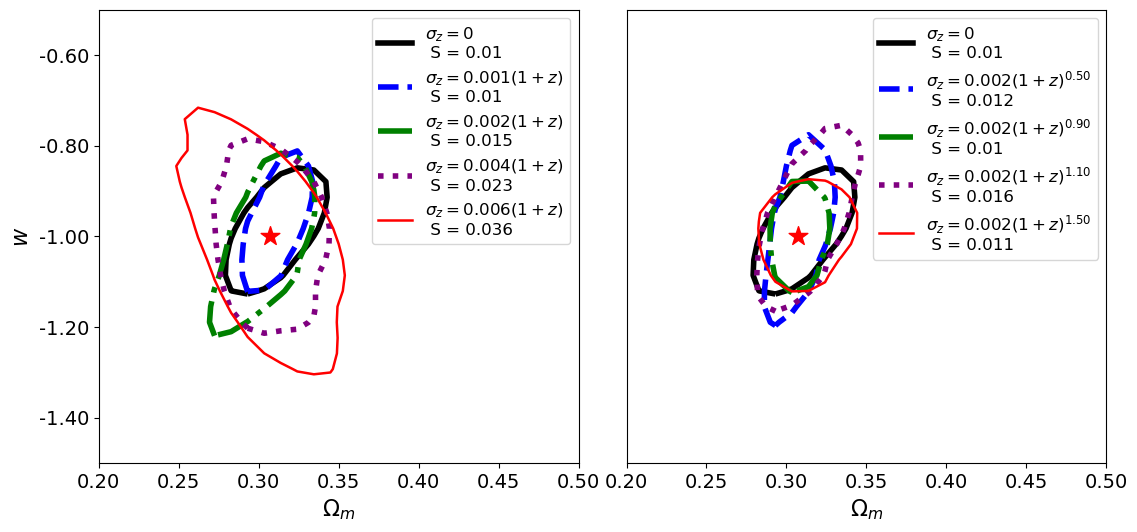}
    \caption{The 68.3\% CL contours in $ \Omega_m$-$w$ space, for different redshift error schemes. The left panel depicts the results for linear error schemes ($\alpha=1$), while the right panel depicts the error schemes with non-linear form ($\alpha\ne 1$). The areas of the 68\% CL contours are calculated and denoted as $S$ in the legend. The contours are produced with grid sampling of $\chi^2$. The grid pixel size and hence the numeric accuracy of $S$ are about $\sim 10^{-3}$.}
    \label{fig:contour}
\end{figure*}

We calculate the $ \chi^2$ values defined by Eq.~\ref{equ:chi2} and~\ref{equ:pz}, and plot the derived 68.3\% confidence level (CL) contours for the $\Omega_m$ and $w$ parameters in Figure~\ref{fig:contour}. The left panel shows the variation of the error amplitude ($\sigma$ parameter), whereas the right panel shows the variation of redshift dependence ($\alpha$ parameter). The areas of the 68\% CL contours, denoted as $S$, are listed for a quantitative comparison. In all cases the redshift errors do not catastrophically worsen the cosmological constraints. In particular, when the amplitude of redshift errors is controled below $\sigma \lesssim 0.002$, the impact of redshift errors is quite mild ($\lesssim 50\%$).

The redshift errors  also alter the direction of degeneracy. For example, while in the $\sigma_z=0$ case there exists a positive correlation between $\Omega_m$ and $w$, in the $\sigma_z=0.006(1+z)$ error scheme the correlation becomes negative. In other words, redshift errors tend to wash out information in some preferred directions. Thus, part of the lost cosmological information might be recovered by doing joint analysis with other cosmological probes.

\section{Concluding Remarks}

For the first time in the literature, we carried out a detailed study of the impact of redshift errors on the tomographic AP analysis. The major findings are: i) the redshift errors typically do not introduce a significant redshift evolution of clustering anisotropy; ii) the LOS smearing typically reduces the RSD contamination and is less harmful to the tomographic AP signal, and therefore often reduces the bias-to-signal ratio; iii) for typical stage-IV slitless spectroscopic surveys, the redshift errors do not catastrophically worsen the cosmological constraints, typically by a factor of $\lesssim 50\%$. 

The present work uses two tomography bins at $z\lesssim 1$ and a phenomenological model $\sigma(1+z)^\alpha$ to account for variable amplitude and running of the redshift errors. More sophisticated models are certainly of interests and are left for future exploration.

Throughout this analysis, we fix the clustering scale of the analysis as $6$-$40 h^{-1}$ Mpc. While this is a reasonable approach for small redshift errors, as suggested by previous works~\citep{LI16,LI18,LI19}, caution might need to be taken for larger redshift uncertainties. This issue is not covered by the present work, and is worthy future investigation.


In summary, our work suggests that the conduction of the tomographic AP analysis is feasible  as long as the error in measuring the galaxy redshifts is not excessively large and does not have excessively non-trivial redshift dependence. We anticipate that the application of the tomographic AP method to stage-IV surveys will yield promising cosmological constraints.

\section*{Acknowledgements}

The CosmoSim database used in this paper is a service by the Leibniz-Institute for Astrophysics Potsdam (AIP).
The MultiDark database was developed in cooperation with 
the Spanish MultiDark Consolider Project CSD2009-00064.

This work is supported by National SKA Program of China No. 2020SKA0110401 and No. 2020SKA0110402; Zhiqi Huang acknowledges the support from the  National Natural Science Foundation of China (NSFC) under Grant No. 12073088; Xiao-Dong Li acknowledges the support from the NSFC grant (No. 11803094), the Science and Technology Program of Guangzhou, China (No. 202002030360), as well as the discussions with Qiyue Qian. We acknowledge the science research grants from the China Manned Space Project with No. CMS-CSST-2021-A03, No. CMS-CSST-2021-B01. 


\section*{Data availability}

The BigMD simulation used in this paper is availalbe via the CosmoSim database (https://www.cosmosim.org/).

\bibliographystyle{mnras}
\bibliography{cites} 








\bsp	
\label{lastpage}
\end{document}